\documentclass[prb,aps,amsmath,amssymb,amsfonts,floatfix,showpacs,twocolumn,superscriptaddress,footinbib]{revtex4-1}

\usepackage[colorlinks=true,citecolor=blue,urlcolor=blue,linkcolor=blue,hyperfigures=true]{hyperref}
\usepackage{graphicx}
\usepackage{amsmath,amsfonts}
\usepackage{xcolor}
\usepackage{color}

\newcommand{\up}{\uparrow}
\newcommand{\dn}{\downarrow}
\newcommand{\kv}{\ensuremath{\mathbf{k}}}
\newcommand{\qv}{\ensuremath{\mathbf{q}}}

\renewcommand{\Im}{\operatorname{Im}}

\begin{document}

\author{Erik G. C. P. van Loon}
\affiliation{Radboud University, Institute for Molecules and Materials, NL-6525 AJ Nijmegen, The Netherlands}

\author{Hartmut Hafermann}
\affiliation{Mathematical and Algorithmic Sciences Lab, Paris Research Center, Huawei Technologies France SASU, 92100 Boulogne Billancourt, France}

\author{Mikhail I. Katsnelson}
\affiliation{Radboud University, Institute for Molecules and Materials, NL-6525 AJ Nijmegen, The Netherlands}


\title{Precursors of the insulating state in the square lattice Hubbard model}

\begin{abstract}
We study the two-dimensional square lattice Hubbard model for small to moderate interaction strengths $1\leq U/t\leq 4$ by means of the ladder dual fermion approach. The non-local correlations beyond dynamical mean-field theory lower the potential energy, lead to a maximum in the uniform susceptibility and induce a pseudogap in the density of states.  
While the self-energy exhibits precursors of a possible insulating phase linked to the appearance of long-range fluctuations, the metallic phase persists within the accessible temperature range. Finite-size effects affect results qualitatively. Upper bounds on the crossover temperature are found to be significantly lower than previously reported dynamical vertex approximation results at $U/t=1$.
\end{abstract}

\maketitle

\section{Introduction}

The Hubbard~\cite{Hubbard63,Gutzwiller63,Kanamori63,Hubbard64} model captures numerous phenomena of strongly correlated electron physics, in particular the Mott metal-insulator transition~\cite{Imada1998}. Despite various efforts, the nature of the change from metal to insulator in the two-dimensional (2D) Hubbard model at half-filling and intermediate to small interaction remains elusive.

At large interaction strength $U$, spin and charge degrees of freedom are decoupled and a gap opens in the density of states already at finite temperature.
The Hubbard model approximately maps to a Heisenberg model in this regime with an effective exchange coupling~\cite{Anderson59} $J = -4t^2/U$, where $t$ is the nearest-neighbor hopping. At zero temperature the local moments order due to the superexchange mechanism. 

At small interaction, the nature of the phase connecting the weakly correlated Fermi liquid at high temperature with the antiferromagnet (AF) at zero temperature is much less clear.  At $U=0$, the Fermi surface is perfectly nested\footnote{We consider the Hubbard model with only nearest-neighbor hopping.} with antiferromagnetic nesting vector $q=(\pi,\pi)$. In one possible scenario, for small finite $U$ a spin-density wave instability develops at $T=0$ and the MIT is a consequence of the backfolding of the Brillouin-zone due to the magnetic ordering. This is called the Slater transition~\cite{Slater51}.
Kyung et al.~\cite{Kyung03} have argued for this scenario based on the two-particle self-consistent approach~\cite{Vilk94,Vilk97,Tremblay12} (TPSC).

In an alternative scenario propagated by Anderson~\cite{anderson97}, the Hubbard model exhibits strong-coupling behavior for both strong \emph{and} weak coupling, so that a Mott gap is present at any finite value of $U$ as in 1D\cite{[{For a discussion of the differences between 1D and 2D, see }] [{}]Boies95}. As the temperature is lowered, local moments develop first because of the MIT and finally order at $T=0$. In this scenario, AF is a consequence of the MIT, contrary to the Slater scenario. Moukouri et al.~\cite{Moukouri01} have argued in favor of this scenario based on the double occupancy and density of states in large cluster DCA calculations. Sch\"afer et al.\cite{Schafer15} identified a $U_c(T)$ based on a downturn in the electronic self-energy at the lowest Matsubara frequencies within the dynamical vertex approximation (D$\Gamma$A) and Quantum Monte Carlo (QMC). This downturn has been interpreted as a destruction of the Fermi surface due to scattering of the electrons at Slater paramagnons, fluctuations with a very large correlation length which can extend over thousands of sites. The downturn in self-energy has been found to correlate with a decrease in potential energy. This is consistent with the cellular DMFT results of Fratino et al.~\cite{Fratino17}, which show a decrease in potential energy when the finite cluster undergoes a transition to the antiferromagnetically ordered state.

In this work, we contribute to the current physical picture by studying the small interaction, low temperature region using the ladder dual fermion approximation (LDFA)~\cite{Rubtsov08,Hafermann2009}. 
The LDFA, as well as D$\Gamma$A belong to a class of methods known as diagrammatic extensions of dynamical mean-field theory~\cite{Rohringer17}. Contrary to cluster methods, they include correlations over length scales covering hundreds of lattice sites. While the self-energy is approximate at any scale, good agreement of the LDFA with benchmarks has been found over a wide parameter range~\cite{Gukelberger2017}.
Despite similarities in the LDFA and D$\Gamma$A, they differ in how they treat the long-range fluctuations which are essential to respect the Mermin-Wagner theorem~\cite{Mermin66}.
Because the physical question discussed here is closely related to the presence of these fluctuations,  we do not necessarily expect the same results in these methods.

In ladder D$\Gamma$A the Mermin-Wagner theorem is a consequence of the so-called Moriya-$\lambda$ correction~\cite{Katanin2009}. Its purpose is to ensure the correct leading asymptotic behavior of the self-energy and can be understood as a mass term in the two-particle propagator (susceptibility) which cuts off the divergence that leads to spurious long-range order.
In the LDFA, the diagrammatic corrections do not alter the leading term in the asymptotic behavior~\footnote{This is because the dual Green function decays as $1/(i\nu)^2$.}. Here the long-range antiferromagnetic fluctuations are included through a self-consistent renormalization procedure which ensures the expected exponential scaling of the susceptibility at low temperature~\cite{Otsuki14}.

In the following section~\ref{sec:method} we discuss the model and some additional properties of the LDFA.
Our numerical results and their implications for the small interaction and low temperature phase diagram of the two-dimensional Hubbard model are discussed in Sec.~\ref{sec:results}.
We conclude in Sec.~\ref{sec:conclusions}.
 
 \section{Model and Method}
 \label{sec:method}
  
The Hubbard model on the 2D square lattice is described by the Hamiltonian
 \begin{align}
 H=
  -t \sum_{\langle ij \rangle} c^\dagger_{j\sigma} c_{i\sigma} + U \sum_i n_{i\up} n_{i\dn}, 
 \end{align}
where $t$ is the nearest-neighbor hopping and our unit of energy. We are interested in small to intermediate coupling $U$ up to half of the bandwidth given by $W=8t$.

The LDFA is a particular diagram resummation scheme in the dual fermion (DF) approach. In DF the lattice model is replaced by a lattice of quantum impurity models which interact via auxiliary, so-called dual fermions. The strong local DMFT-like correlations are treated at the level of the impurity model, while non-local correlations are included diagrammatically. For a specific choice of self-consistency condition on the impurities, DMFT corresponds to non-interacting dual fermions~\cite{Rubtsov08}. 
The latter couple to the physical fermions of the same flavor locally. Diagrams in terms of dual fermions can therefore be constructed based on physical considerations. In particular, we expect long-range particle-hole fluctuations to be dominant. The corresponding LDFA self-energy has the form
\begin{align}
\widetilde{\Sigma}_{\mathbf{k}\nu} =& -\frac{1}{2}\frac{T^{2}}{N^{2}}\sum_{k\, q}\sum_{r} A_{r}F_{r}^{\nu\nu'\omega}\widetilde{G}_{\mathbf{k}'\nu'}\widetilde{G}_{\mathbf{k}'+\qv,\nu'+\omega}\widetilde{G}_{\mathbf{k}+\qv,\nu+\omega}\notag\\
&\times[\widetilde{F}_{\text{lad},r,\qv}^{\nu\nu'\omega}-\frac{1}{2}F_{r}^{\nu\nu'\omega}],
\label{eq:sigmaldfa}
\end{align}
where the ladder diagrams are generated by the Bethe-Salpeter equation (BSE) for the dual vertex $\widetilde{F}_{\text{lad},r,\qv}^{\nu\nu'\omega}$ of the lattice in the particle-hole channel,
\begin{align}
\widetilde{F}_{\text{lad},r,\qv}^{\nu\nu'\omega} = F_{r}^{\nu\nu'\omega}\! -\!\frac{T}{N}\sum_{k} F_{r}^{\nu\nu''\omega}\widetilde{G}_{\mathbf{k}''\nu''}\widetilde{G}_{\mathbf{k}''+\qv,\nu''+\omega}\widetilde{F}_{\text{lad},r,\qv}^{\nu''\nu'\omega}.
\label{eq:bse}
\end{align}
Here we have introduced sums over four-momenta $k=(\mathbf{k},\nu)$, $q=(\mathbf{q},\omega)$ and the spin and charge channels $r=\text{sp},\text{ch}$. $\nu$ and $\omega$ denote the discrete fermionic and bosonic Matsubara frequencies, respectively.
We further have $A_{\text{ch}}=1$, $A_{\text{sp}}=3$, where the latter accounts for the degeneracy of the spin bosonic excitations. $T$ denotes temperature and $N$ is the total number of lattice sites.
The second term in angular brackets avoids over-counting of the second-order diagram.
$F_{r}^{\nu\nu'\omega}$ is the exact local reducible vertex of the impurity model. If we approximate the lattice vertex by the local one, $\widetilde{F}_{\text{lad},r,\qv}^{\nu\nu'\omega}\approx F_{r}^{\nu\nu'\omega}$, we obtain the second-order approximation~\cite{Rubtsov08,Rubtsov09} DF$^{(2)}$.
Note that the LDFA includes diagrams from both the horizontal and vertical particle-hole channels (see for example the discussion in Ref.~\onlinecite{Gukelberger2017}). In the spin channel, these diagrams generate the collective paramagnon excitations~\cite{Wilhelm2015}. Remarkably, the LDFA reproduces non-mean-field critical exponents~\cite{Antipov14,Hirschmeier15}.
We refer the reader to Refs.~\onlinecite{Rohringer17,Rubtsov08} for further details on the method.
 
Below the DMFT N\'eel temperature $T_{\text{N}}^{\text{DMFT}}$ we have to include the long-range fluctuations which destroy the AF order of the underlying mean-field. To this end, the dual Green's functions $\tilde{G}_{\kv\nu}$ in above equations are self-consistently renormalized: The self-energy $\tilde{\Sigma}_{\mathbf{k}\nu}$ is calculated starting from an initial guess for the dual Green's function (typically the bare dual Green's function). A new dual Green's function is obtained via Dyson's equation, which in turn is inserted into \eqref{eq:bse} and \eqref{eq:sigmaldfa} to calculate a new self-energy (the impurity vertex is fixed). This process is repeated until self-consistency. 

Since below $T_{\text{N}}^{\text{DMFT}}$ the BSE~\eqref{eq:bse} initially diverges, we cut off the AF fluctuations in the initial iterations of this inner self-consistency by restricting the eigenvalues of the BSE to values strictly smaller than 1. Once the iterations converge the cutoff is removed. When all BSE eigenvalues are smaller than unity in the final iteration the solution is well-defined and independent of the cutoff~\cite{Otsuki14}. The scheme is not guaranteed to converge and the number of iterations may diverge, which ultimately limits the accessible temperature range.
The impurity model hybridization function is adjusted in an \emph{outer} self-consistency loop based on the condition that the lowest-order dual diagram vanishes~\cite{Rubtsov08}. At the values of $U$ we are interested in, its effect is merely a small enhancement of the imaginary part of the impurity self-energy.

The calculations are carried out on lattices of finite size subject to periodic boundary conditions imposed by the discrete Fourier transform. 
Once the calculation is converged, we compute the physical self-energy $\Sigma_{\kv\nu}$ and momentum-resolved susceptibility~\cite{Brener08}, $\chi(\omega,\qv) = \langle \hat{S}^z \hat{S}^z \rangle_{\omega,\qv}$ with $S^z = (\hat{n}_\up-\hat{n}_\dn)/2$.
 
 \section{Results}
 \label{sec:results} 
 
 \subsection{Energetics}
 
 \begin{figure}[t]
  \includegraphics[]{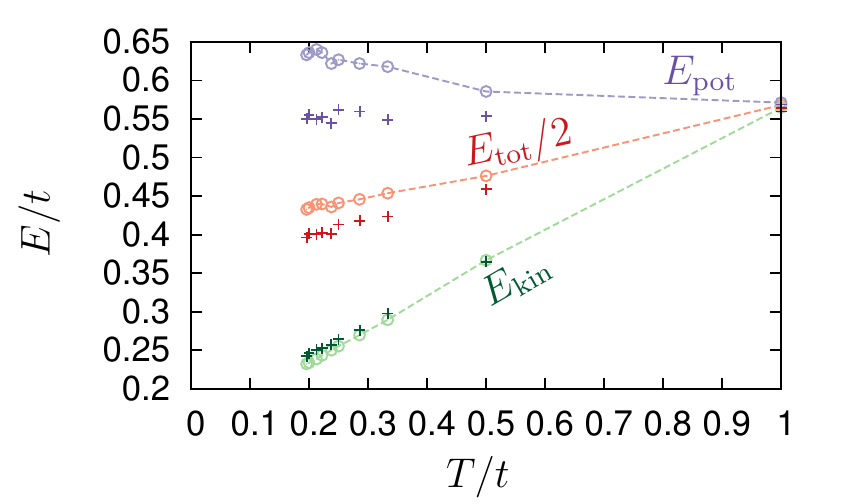}
  \caption{Energetics at $U/t=4$ obtained on a $64\times 64$ lattice. The kinetic energy is measured with respect to the kinetic energy at $T=0$ and $U=0$, $E^0_\text{kin}=-16/\pi^2$. Dashed lines and circles are the DMFT results, plus signs are LDFA.}
  \label{fig:energeticsU4}
 \end{figure} 
 
 \begin{figure}[t]
  \includegraphics[]{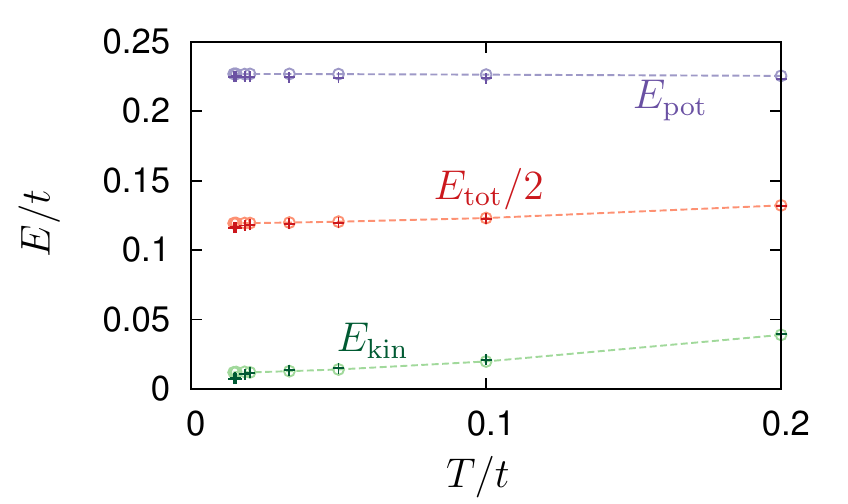}
  \caption{Energetics at $U/t=1$ obtained on a $128\times 128$ lattice. The kinetic energy is measured with respect to the kinetic energy at $T=0$ and $U=0$, $E^0_\text{kin}=-16/\pi^2$. Dashed lines and circles are the DMFT results, plus signs are LDFA.}
  \label{fig:energeticsU1}
 \end{figure}  

We first discuss the energetics of the model. Entropy disfavors antiferromagnetism, so the AF transition is driven by the interplay of kinetic and potential energy. In the picture of nearly free electrons at small interaction, AF ordering reduces the double occupancy.
Magnetic ordering is hence potential energy driven. This is the Slater regime.
At strong interaction on the other hand, double occupancy is largely suppressed. AF ordering promotes hopping processes and the (negative) kinetic energy is lowered. Magnetic ordering is stabilized through the reduction in kinetic energy. This is the Heisenberg regime. 
It is  true even in the Mott insulator, where the exchange coupling $J=-4t^{2}/U$ is mediated through virtual hopping processes.

In DMFT it is possible to compute the energy difference between the ordered and unordered states. In the Slater regime, the ordered state has a lower potential energy, but higher kinetic energy compared to the unordered state at $T=0$~\cite{Taranto12}. This implies that indeed the potential energy stabilizes the ordered state. 
At large $U$, in the Heisenberg regime, the situation is opposite. 
These conclusions remain true when short-range non-local correlations come into play.
Both $2\times 2$ CDMFT~\cite{Fratino17} and plaquette DCA~\cite{Gull08} calculations in the Slater regime show that the potential energy is lowered compared to DMFT. This is expected because the DCA includes antiferromagnetic correlations. 
At large $U$, on the other hand, the four-site plaquette has a higher potential energy. 
Fratino et al.~\cite{Fratino17} further argue that the interaction scale at which the system switches from Slater- to Heisenberg behavior is given by the critical $U$ of the underlying normal-state Mott transition of the plaquette.

D$\Gamma$A includes nonlocal correlations diagrammatically and up to significantly larger length scales compared to the cluster calculations. The correlations increase the kinetic energy at small $U$, while at large $U$ they decrease it~\cite{Rohringer16}. The potential energy was found to be reduced compared to DMFT at all studied values of $U$, both in the Slater and in the Heisenberg regime. However, the ambiguity of the potential energy in DMFT complicates the analysis~\cite{vanLoon16,Rohringer16}.

In Figs.~\ref{fig:energeticsU4} and~\ref{fig:energeticsU1} we show the energetics extracted from LDFA calculations as a function of temperature at $U/t=4$ and $U/t=1$, respectively. In both cases we find that the potential energy is lowered in LDFA compared to DMFT (albeit only sightly at $U/t=1$) in accordance with the cluster DMFT and D$\Gamma$A results and as expected in the Slater regime. At $U/t=4$ the kinetic energy is also somewhat higher than in DMFT, while it seems slightly lower at $U/t=1$ and low temperature. Here relatively large uncertainties~\footnote{The LDFA energetics are obtained through frequency- and momentum summations of  single-particle quantities as in~\onlinecite{Rohringer16} including tail corrections, which are susceptible to numerical errors. In DMFT, the kinetic energy is obtained from the average perturbation order in the CTHYB solver~\cite{Haule07} and the potential energy from the Migdal-Galitskii formula~\cite{Krien17}. The Matsubara frequency summations are most difficult at low temperatures, where very many frequencies are needed.
}
however prevent a definite statement (note the scale compared to $U/t=4$).
The total energy of the system is lower when nonlocal correlations are included.

 \subsection{Density of States}
 \label{sec:dos}
 \begin{figure}[!t]
  \includegraphics[width=\columnwidth,angle=0]{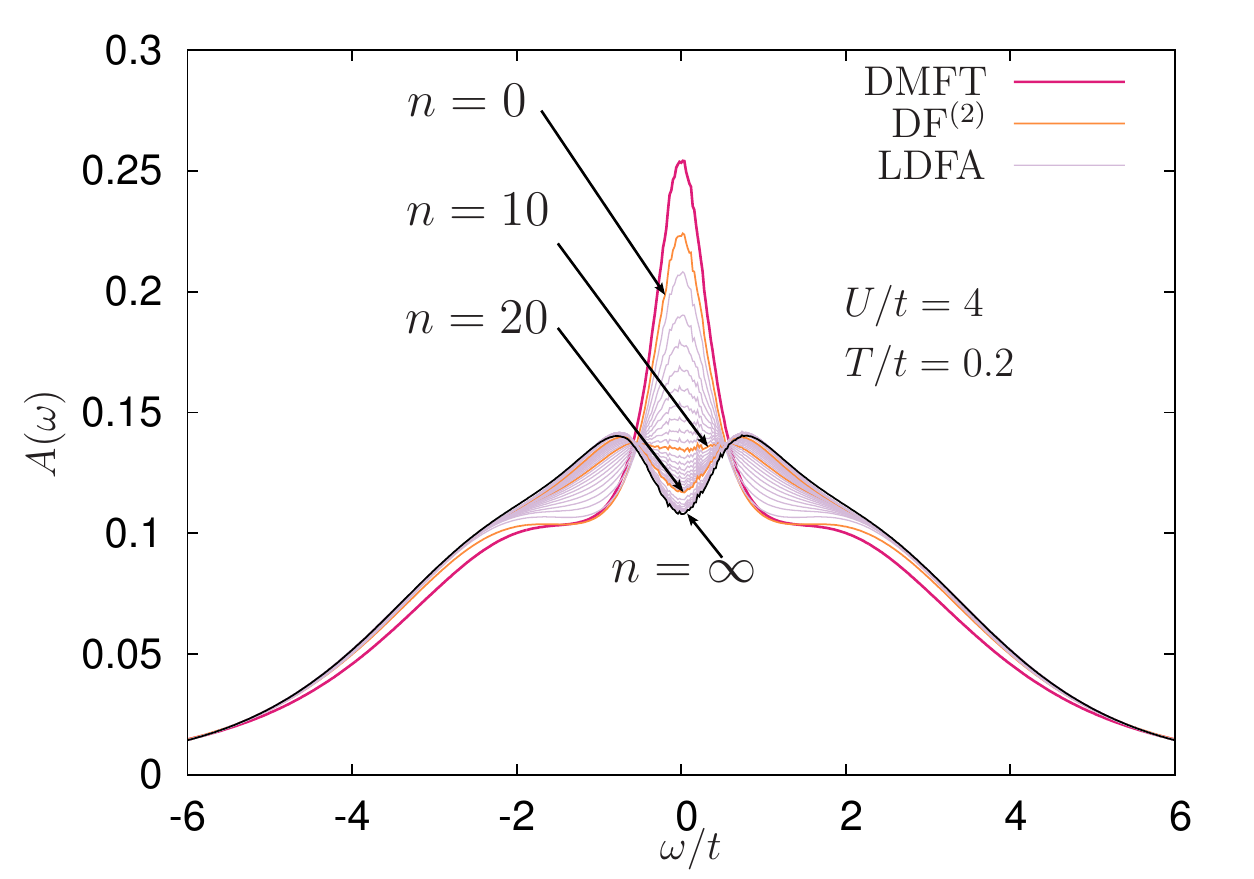}
  \caption{Maximum entropy local density of states obtained within DMFT and ladder DF. For a given $n$, the self-energy includes ladder diagrams up to order $n+2$ in the local vertex. The case $n=0$ corresponds to the second-order approximation DF$^{(2)}$.}
  \label{fig:pseudogapU4}
 \end{figure} 
The non-local fluctuations have a drastic effect on the local density of states (DOS) as seen in Fig.~\ref{fig:pseudogapU4}.
While the DMFT solution exhibits a quasiparticle peak, the spectral weight is reduced in DF. The reduction is small in second-order DF (labeled DF$^{(2)})$, but increases with order of the ladder diagrams. Remarkably, diagrams at all orders contribute to the pseudogap.
The dual Green's function decays rather rapidly in real space, with exponential decay on a length scale~\footnote{This scale is determined by an exponential fit to $G(x,y,\tau=\beta/2)$ along the diagonal $x=y$. The decay is highly anisotropic and the decay length is longest along the diagonal.} $\xi_{G}\approx 2.8$. Low-order diagrams hence mediate short-range correlations, while long-range correlations require high diagram orders. They nevertheless contribute to short-range correlations as well.

The pseudogap develops in a regime where magnetic fluctuations are very strong and is linked to a reduction in potential energy, in accordance with DCA~\cite{Gull08}. The temperature here is slightly below the DMFT N\'eel temperature. The AF correlation length is of the order of $\xi_{\text{AF}}\sim 4$ (see Sec.~\ref{sec:finsize}).
Extended AF fluctuations are hence present in the system and mediate correlations between electrons. The correlations may be singlet-like, which can lead to a strong suppression of the density of states and ultimately open the gap~\cite{Hafermann2009a,Gull08}. These results are consistent with those of Ref.~\onlinecite{Moukouri01}, where a gap fully opens at $U/t=4$ and $T/t=0.125$.  Unfortunately we cannot reach such low temperatures at this value of $U$.
On the other hand, the opening of the gap can be interpreted as a consequence of the scattering of quasiparticles off the AF fluctuations, which we include by construction in the LDFA. This interpretation and our numerical results are consistent with the D$\Gamma$A study of Ref.~\onlinecite{Schafer15}, which pointed out the importance of Slater paramagnons. The qualitative DOS predicted by TPSC at crossover temperatures~\cite{Vilk97} is consistent with the LDFA results.

\begin{figure}[!t]
\includegraphics[]{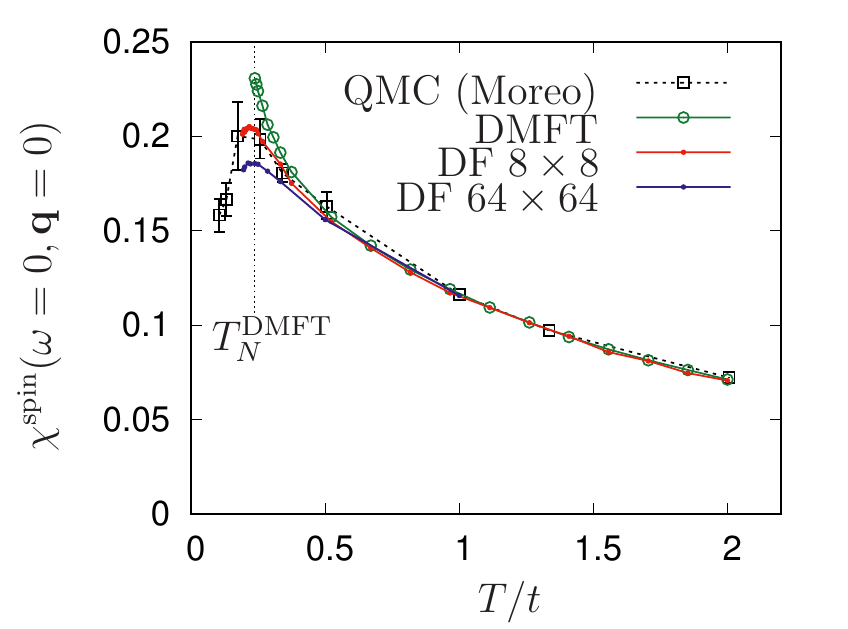}
\caption{Uniform spin susceptibility as a function of temperature compute with different methods, all at $U/t=4$. In QMC and LDFA, $\chi^{\text{sp}}$ exhibits a maximum at the effective exchange energy scale. The QMC results are taken from Ref.~\onlinecite{Moreo93} for an $8\times 8$ lattice.}
\label{fig:suscU4}
 \end{figure}

 \subsection{Uniform susceptibility}
 
Figure~\ref{fig:suscU4} shows the temperature dependence of the uniform spin susceptibility $\chi^{\text{sp}}$ at $U/t=4$. 
We find a decreasing susceptibility at high temperatures, even though Curie's law~\cite{Moriya85} $\chi \propto 1/T$ only sets in at higher temperatures, $T\approx W=8t$.
While the results of different methods agree at high temperature, they are qualitatively different at low $T$. In DMFT, $\chi$ continues to increase up to the point where the antiferromagnetic susceptibility diverges due to the second-order transition to the mean-field antiferromagnetic state at $T_{\text{N}}^{\text{DMFT}}$ (not shown).
In LDFA the susceptibility decreases at low temperature and exhibits a maximum. 
AF correlations that build up reduce the uniform susceptibility. This occurs at the energy scale of the effective exchange interaction $J$ between neighboring sites.
DMFT does not include such nonlocal correlations, so that the maximum is absent.
Note that the maximum occurs at a slightly lower temperature than $T_{\text{N}}^{\text{DMFT}}$. At $T_{\text{N}}^{\text{DMFT}}$ the ladder diagram series~\eqref{eq:bse} diverges, which in turn causes large effects in the self-consistent renormalization of the Green's functions. The result are strong magnetic fluctuations which destroy the mean-field long-range order in the underlying DMFT solution. At the same time they lead to spin correlations between sites.

The LDFA susceptibility is in excellent agreement with lattice QMC results~\cite{Moreo93} for the same lattice size. The comparison with results for a larger lattice reveals that finite size effects play a role. The magnitude of the susceptibility is significantly reduced (by about 10\%) in the larger system. The position of the maximum however is not affected by finite size effects, which is consistent with its interpretation as an effective exchange energy scale.

\begin{figure}
\includegraphics[]{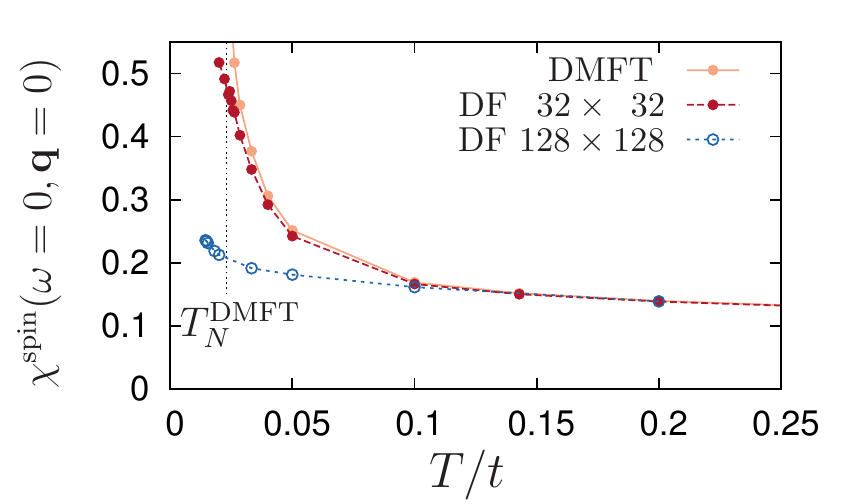}
\caption{Uniform spin susceptibility as a function of temperature, at $U/t=1$.}
\label{fig:suscU1}
 \end{figure}  

\begin{figure}[b]
\includegraphics[]{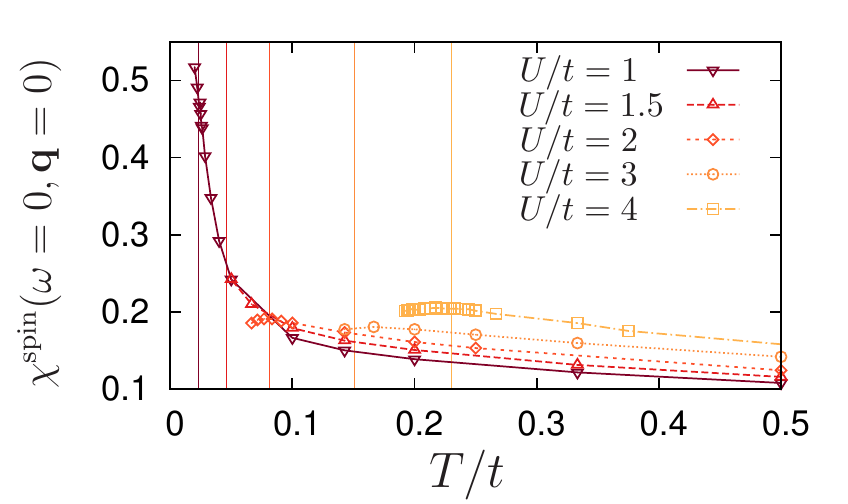}
\caption{Uniform spin susceptibility as a function of temperature, for various values of $U$. Note that all simulations have been performed on a $32\times 32$ lattice, except for the $U/t=4$ simulations which were performed on a $64\times 64$ lattice. The vertical lines show the respective $T_N^\text{DMFT}$.}
\label{fig:suscUs}
 \end{figure}  

The uniform spin susceptibility for $U/t=1$ is shown in Fig.~\ref{fig:suscU1}. As for $U/t=4$, the nonlocal correlations reduce the susceptibility compared to DMFT, especially at lower temperatures. In LDFA, the magnitude of the susceptibility is reduced on the larger lattice, even more strongly than for $U/t=4$. However we do not find a maximum in the susceptibility at the accessible temperatures down to $T/t=1/70\approx 0.014$. 
At half-filling and $U=0$, there is a Van Hove singularity in the electron density of states exactly at the Fermi level and the uniform susceptibility shows a logarithmic divergence as $T\rightarrow 0$. Based on our results, it is not possible to distinguish between a maximum at a finite but very small temperature, or at $T=0$.
The DMFT N\'eel temperature is between $T/t=0.025$ and $T/t=0.022$, showing that it is not necessarily linked to the occurrence of the maximum as for $U/t=4$.

In Fig.~\ref{fig:suscUs} we show results for several values of $U$. The location of the maxima for $U/t=2, 3$ and $4$ agree within error bars with the QMC results of Ref.~\onlinecite{Paiva10}, where the location of the maximum has been used to find the energy scale at which magnetic correlations become important in optical lattice experiments~\footnote{Note that the results are not comparable in magnitude. The QMC calculations in Ref.~\onlinecite{Paiva10} we performed on a $10\times 10$ lattice for which we expect strong finite-size effects. Note also that the definitions of the susceptibilities differ by a factor of 4.}.
The location of the maximum in the susceptibility decreases with decreasing $U$, showing that the physics is clearly not Heisenberg-like. For the $U$ values where we obtain a maximum, it is close to the DMFT N\'eel temperature.
 
 \subsection{Finite-size effects}
 \label{sec:finsize}

 \begin{figure}[t]
\includegraphics[]{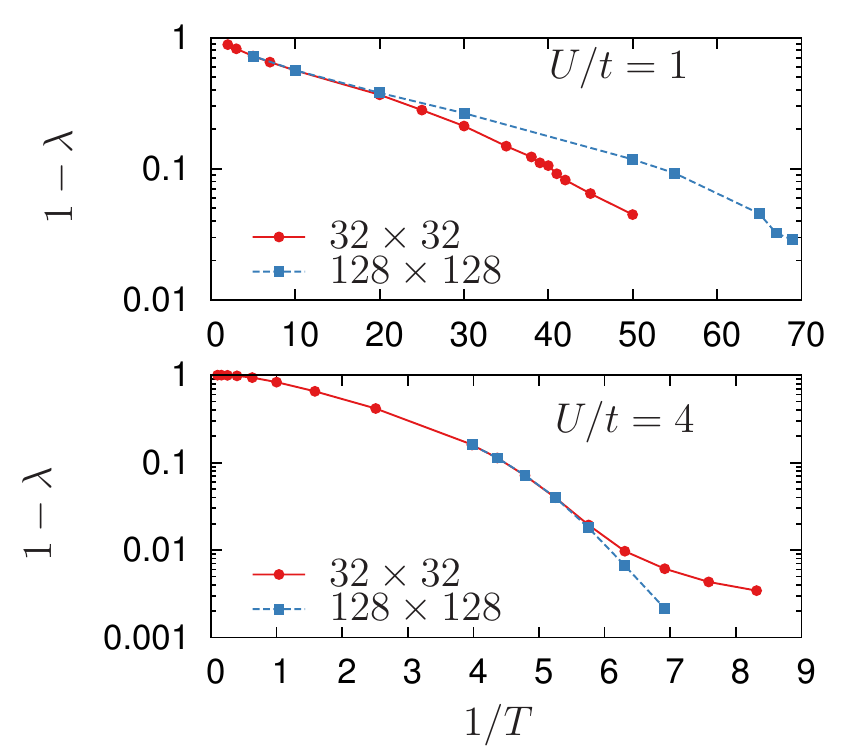}
\caption{Leading LDFA antiferromagnetic eigenvalue as a function of temperature. Note the different x-axes for $U/t=1$ and $U/t=4$. The results for $U/t=4$ are reproduced from Ref.~\onlinecite{Otsuki14}.}
\label{fig:ev}
\end{figure}  

Figure~\ref{fig:ev} shows the leading eigenvalue of the BSE at the AF wave vector for $U/t=1$ and $U/t=4$ respectively, for different lattice sizes.
At low temperature the antiferromagnetic susceptibility is expected to exhibit an exponential scaling $\chi_{\text{AF}}\sim e^{\Delta/T}$~\cite{Hasenfratz1991,Otsuki14}. It follows that the leading eigenvalue behaves as $(\chi_{\text{AF}})^{-1}\sim 1-\lambda\sim\exp(-\Delta/T)$. $1-\lambda$ remains non-zero for finite $T$ as required by the Mermin-Wagner theorem. 
We observe the scaling for $U/t=4$ for the larger lattice, and not at all for $U/t=1$.

\begin{figure}[t]
  \includegraphics[]{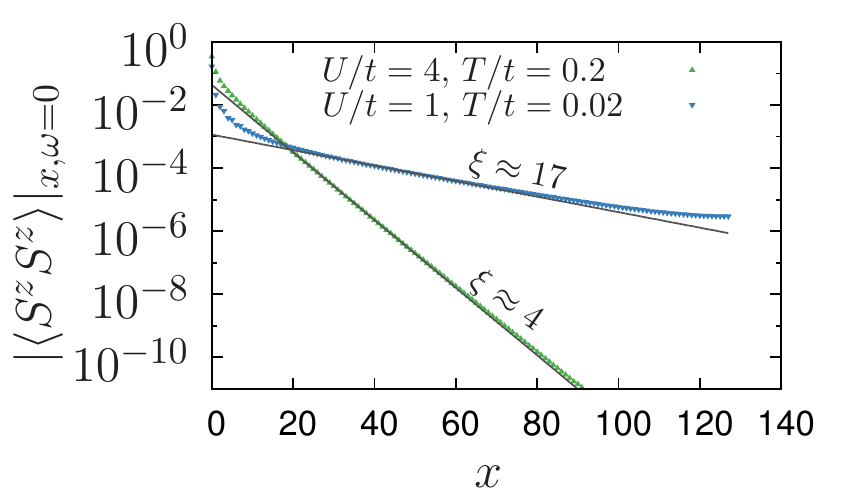}
  \caption{Spin correlation function in real space on a $256\times 256$ lattice along the line $(x,0)$, where the coordinate $x$ is given in units of the lattice constant. The grey lines show an exponential fit with correlation length $\xi$. Finite-size become apparent at distance of roughly half the linear latte size.
}
  \label{fig:correl}
 \end{figure}

Figure~\ref{fig:correl} shows the spin correlation function in real space. At $U/t=4$ and $T/t=0.2$ the AF correlation length~\footnote{This correlation length is determined by fitting $a \exp(-x/\xi)$ to $\chi(x,y,\omega=0)$ along the line $y=0$. The fit is performed at intermediate values of $x$, for $x \approx L$ the periodicity of the lattice is noticeable and leads to deviations from exponential decay.} is short, $\xi_{\text{AF}}\approx 4$. This is consistent with the fact that no finite-size effects are visible at this temperature in Fig.~\ref{fig:ev}. At $U/t=1$ and $T/t=0.02$, the correlation length is significantly longer, $\xi_{\text{AF}}\approx 17$, but unlikely to explain the absence of the scaling for the $128\times 128$ lattice.
 
In addition to the length scale associated with two-particle fluctuations, we can introduce a scale related to single-particle properties obtained from the Green's function at $\tau=\beta/2$, $\xi_{G}$.
For $U/t=4$ and $T/t=0.2$, we find $\xi_{G} \approx 2.8 < \xi_{\text{AF}} \approx 4$, while for $U/t=1$ and $T/t=0.02$ we obtain $\xi_{G}\approx 35$, which is larger than $\xi_{\text{AF}}$ and the smaller lattice and may explain why finite-size effects act in opposite directions for $U=4$ and $U=1$. Even though we cannot rule out that the exponential scaling of $1-\lambda$ is obscured by finite size effects, we expect that it sets in at even lower, inaccessible temperatures.

\begin{figure}[b]
  \includegraphics{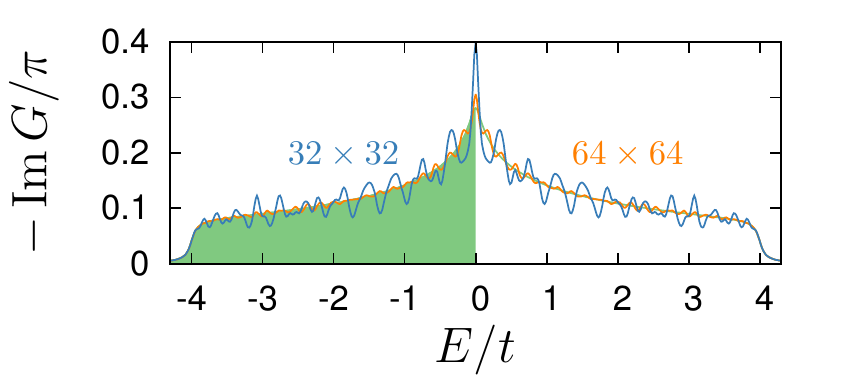}
  \caption{Noninteracting density of states $-\frac{1}{\pi}\Im G_0(E+i\eta)$ at $T/ t=0.02$, with broadening $\eta=\pi T$, determined using three different lattice sizes. The filled, green curve uses a $256\times 256$ lattice, the lines use smaller lattices. Significant finite-size effects are visible for $N_x<64$.
  }
  \label{fig:noninteracting_dos}
 \end{figure}

\begin{figure}[t]
  \includegraphics{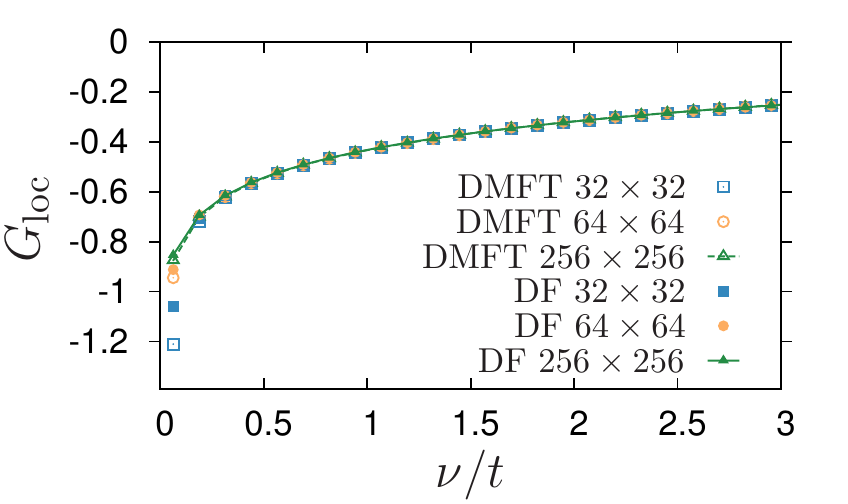}
  \caption{Comparison of the local physics according to DMFT and LDFA, at $U/t=1$ and $T/ t = 0.02$ and three different lattice sizes. There are significant finite-size effects both in DMFT and in LDFA separately, while both  give similar results for the same lattice size.}
  \label{fig:U1:DFvsDMFT}
 \end{figure}

\begin{figure}[t]
  \includegraphics{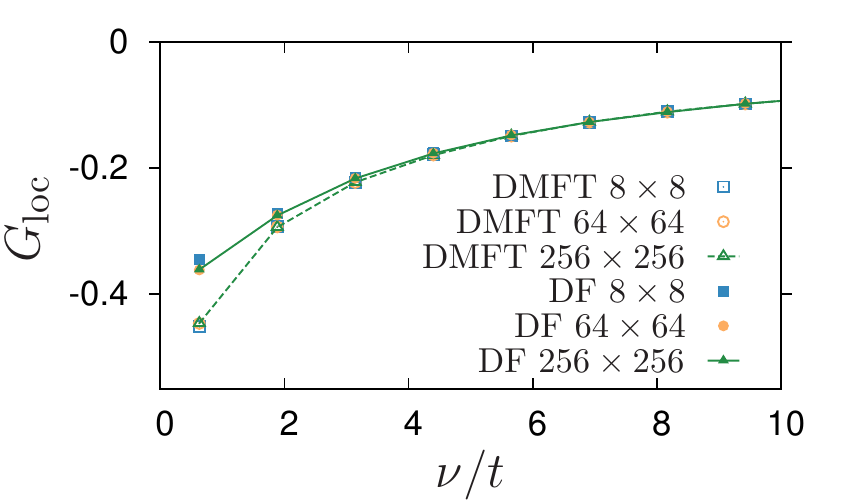}
  \caption{Comparison of the local Green's function according to DMFT and LDFA, at $U/t=4$ and $T/ t = 0.2$ and three different lattice sizes.}
  \label{fig:U4:DFvsDMFT}
 \end{figure}

Figs.~\ref{fig:U1:DFvsDMFT} and \ref{fig:U4:DFvsDMFT} show the local Green's function for the same $U$ values. For $U/t=4$, finite-size effects are clearly absent for large lattices. For $U/t=1$ small finite-size effects are visible even on the largest lattice. Note that the finite-size effects are also visible in DMFT. This underlines that they are related to the momentum discretization. 
Fig.~\ref{fig:noninteracting_dos} illustrates this for the noninteracting DOS, which requires of the order of $64\times 64$ points at $T=0.02$ to be accurately represented.

 \subsection{Self-energy}

\begin{figure}[t!]
  \includegraphics{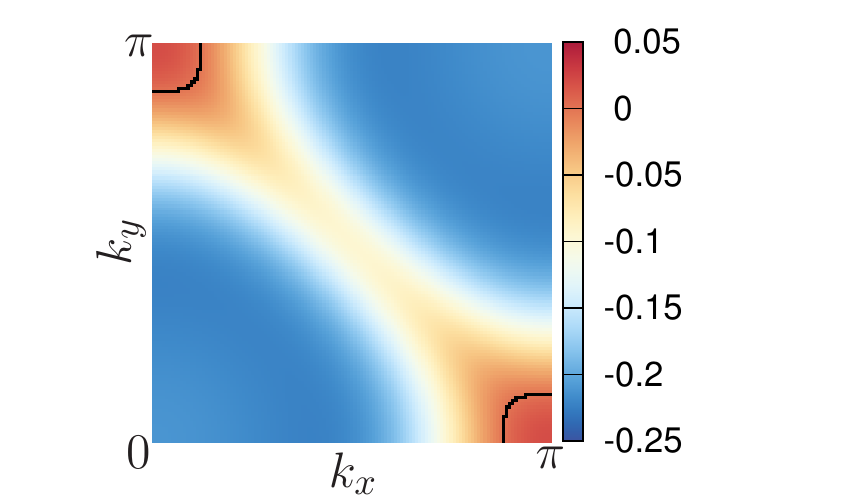}
  \caption{
  Difference in self-energy at the lowest two Matsubara frequencies, $\Im \Sigma_{\nu_1}-\Im \Sigma_{\nu_0}$, at $U/t=4$ and $T/t=0.2$ on a $256\times 256$ lattice. Positive values indicate a downturn.
  Only a quarter of the Brillouin Zone is shown. The black contours indicate where the self-energy difference is zero.
  }
  \label{fig:U4:sigmamap}
 \end{figure}

In Fig.~\ref{fig:U4:sigmamap}, we plot the momentum-resolved difference of the self-energy at the two lowest Matsubara frequencies at $U/t=4$ and $T/t=0.2$. The presence of positive values near the point $\mathbf{k}=(\pi,0)$~\footnote{This is consistent with the observation that the impact of spin fluctuations on the self-energy is largest at the Van Hove singularity~\cite{Vilk97}.} indicates a downturn in the self-energy and signals a breakdown of Fermi liquid theory.
This feature is robust with respect to the lattice size and consistent with the presence of the pseudogap.
At the studied temperature, the downturn only occurs close to $\mathbf{k}=(\pi,0)$. By lowering the temperature, the momentum space region with positive values of the self-energy difference grows. D$\Gamma$A results indicate that there should be a downturn along the entire Fermi surface at $T/t\approx 0.11$ which is outside the accesible temperature range $T/t\geq 0.196$.
We do not find a divergence of the self-energy and hence no opening of a gap at finite temperature. 

\begin{figure}[t!]
 \includegraphics{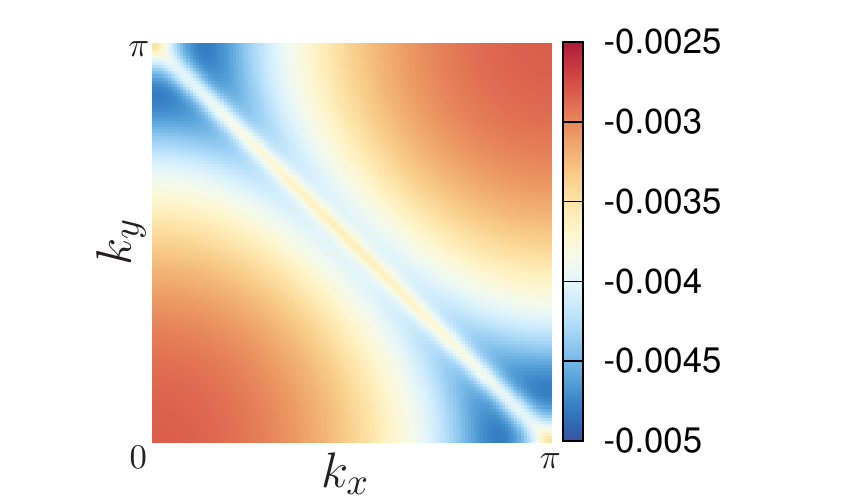}
 \caption{\label{fig:U1_BETA50_Sigmamap} Difference in self-energy at the lowest two Matsubara frequencies, $\Im \Sigma_{\nu_1}-\Im \Sigma_{\nu_0}$ at $U/t=1$, $T/ t = 0.02$, on a $256\times 256$ lattice. Note the extremely sharp the feature at the Fermi surface. However, it does not become positive, which would signal a downturn.}
\end{figure}

In Fig.~\ref{fig:U1_BETA50_Sigmamap}, we show corresponding results for $U/t=1$ and $T/t=0.02$. We do not observe a downturn, but values close to zero (a flattening of the self-energy at low frequencies) in a very narrow strip along the Fermi surface (the diagonal from top left to bottom right). Compared to Fig.~\ref{fig:U4:sigmamap}, the tendency towards a downturn occurs in a much narrower part of the Brillouin Zone. This suggests that the phenomenon is related to AF fluctuations on very long length scales.

\begin{figure}[b]
 \includegraphics{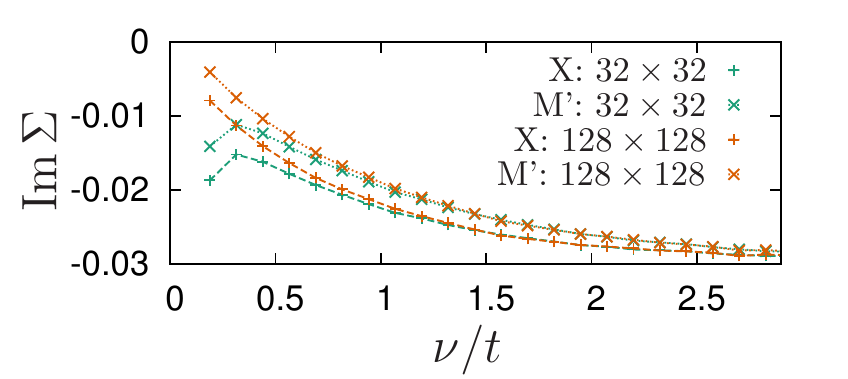}
 \caption{\label{fig:finitesizeSigma} Finite-size effects in the self-energy, at $U/t=1$ and $T/t=0.02$. The self-energy is shown at the X$=(\pi,0)$ and M'$=(\pi/2,\pi/2)$ points.}
 \end{figure}

Figure~\ref{fig:finitesizeSigma} shows that finite-size effects can change behavior qualitatively. A downturn in the self-energy occurs on a $32\times 32$ lattice, while it is absent on the larger $128\times 128$ lattice. 
If the scattering of electrons at magnetic fluctuations induces the downturn, it is conceivable that the downturn occurs because the antiferromagnetic susceptibility is overestimated, as visible in Figs.~\ref{fig:suscU1} and \ref{fig:ev}. The downturn, and ultimately an opening of the gap may still appear at lower temperatures. However we do not observe a downturn in the accessible temperature range. This excludes a crossover down to $T/t=1/69$, in contrast to the crossover temperature of $T/t \approx 1/38$ obtained in D$\Gamma$A~\cite{Schafer15}. The discrepancy between both methods may lie in the way they treat the magnetic fluctuations.

Weak-coupling approximations can put these results into perspective. The Hartree-Fock gap~\cite{Hirsch85,Borejsza04} $\Delta E = 32t \exp(-2\pi \sqrt{t/U})$ is exponentially small in $U$, and predicts a typical energy scale $0.06 t$ at $U/t=1$ due to the significant prefactor. The temperatures studied here are actually below this scale. Renormalization-group analysis~\cite{Halboth00}, on the other hand, suggests a critical energy scale of approximately $0.02t$ for the formation of bound particle-hole pairs, which is exactly in the range studied here.
 
 \section{Conclusions}
 \label{sec:conclusions}

We have studied the half-filled 2D Hubbard model on the square lattice in the interaction range $1\leq U/t\leq 4$ using the LDFA. In accordance with D$\Gamma$A~\cite{Rohringer16} and cellular DMFT~\cite{Fratino17,Gull08} results, non-local AF correlations reduce the potential energy as expected in the Slater regime. Strong AF fluctuations develop in the vicinity of the DMFT N\'eel temperature. 
Scattering of electrons off these fluctuations leads to (presumably singlet-like) correlations and to a pseudogap in the density of states. A concomitant downturn is observed in the self-energy at $U/t=4$. We further find a maximum in the uniform susceptibility in good agreement to Monte Carlo results. It can be explained through the buildup of AF correlations on an effective exchange energy scale $J$. This scale gets smaller when $U$ decreases.

We found that finite-size effects can alter the results qualitatively. While they play a minor role at $U/t=4$ due to relatively short correlations lengths, they lead to a spurious downturn in the self-energy at $U/t=1$.
Nevertheless a tendency to a downturn in the self-energy is observed for $U/t=1$ in a very narrow region in the Brillouin zone along the Fermi surface, which suggests that a possible crossover would be associated with long-range AF fluctuations.

We do not find a crossover to an insulator in the accessible temperature range for $1\leq U/t\leq 4$.
From our calculations we obtain an upper bound $T_\text{downturn}/t < 0.0145$ on the crossover temperature for $U/t=1$, which is significantly lower than the value reported in Ref.~\onlinecite{Schafer15}. The origin of this discrepancy between LDFA and D$\Gamma$A  might lie in the way the methods treat the mean-field divergence of the AF susceptibility.

\acknowledgments
 
We thank Emanuel Gull, Georg Rohringer, Thomas Sch\"afer  and  Alessandro Toschi for useful discussions. E.G.C.P. v. L. and M.I.K. acknowledge support from ERC Advanced Grant 338957 FEMTO/NANO.
Our implementation is based on the ALPS~\cite{ALPS2} framework and on the impurity solver of Ref.~\onlinecite{Hafermann13}.

 \bibliography{references}

\end{document}